\documentstyle[aps,prl,preprint,floats,epsf]{revtex} %For preprints, drafts
 
\begin{document}
\preprint{\tighten\vbox{\hbox{\hfil CLEO CONF 99-14}
}}

%\tighten

\title{Study of charmless hadronic B decays into the final states
$K\pi, \pi\pi$, and $KK$, with the first observation of  $B \to \pi^+ \pi^-$
and $B\to K^0\pi^0$}
%CONTACT PERSONS: Jim Alexander (jima@mail.lns.cornell.edu)
%		  Peter Gaidarev (pg@lns62.lns.cornell.edu)
%                 Frank Wuerthwein (fkw@lns62.lns.cornell.edu)

\author{CLEO Collaboration}
\date{\today}
\maketitle
\tighten

\begin{abstract} 
% Insert abstract here.
We have studied charmless hadronic decays of $B$ mesons into
two-body final states with kaons and pions.  We present preliminary results
based on 9.66 million $B\bar{B}$~pairs collected with the CLEO
detector.  We have made the first observation of the decay
$B \rightarrow \pi^+\pi^-$, with the branching fraction of $Br(B
\rightarrow \pi^+ \pi^-) = (4.7^{+1.8}_{-1.5} \pm 0.6) \times
10^{-6}$. We have also observed for the first time the
decay $B \rightarrow K^0\pi^0$ with
the branching fraction of $Br(B \rightarrow K^0 \pi^0) =
(14.8^{+5.9+2.4}_{-5.1-3.3}) \times 10^{-6}$, thus completing the set
of four $K\pi$ branching fraction measurements.  We present
improved measurements for the decays $B\to K^{\pm}\pi^{\mp}$,
$B^{\pm}\to K^0\pi^{\pm}$, and $B^{\pm}\to K^{\pm}\pi^0$. We use
these and other charmless hadronic B decays to make a first
determination of
the value of the weak phase 
${\rm Arg}(V^*_{ub})=\gamma = {113^\circ}^{+25^\circ}_{-23^\circ}$.

\end{abstract}

\newpage
{
\renewcommand{\thefootnote}{\fnsymbol{footnote}}

% Insert author and address list here
\begin{center}
Y.~Kwon,$^{1,}$%
\footnote{Permanent address: Yonsei University, Seoul 120-749, Korea.}
A.L.~Lyon,$^{1}$ E.~H.~Thorndike,$^{1}$
C.~P.~Jessop,$^{2}$ K.~Lingel,$^{2}$ H.~Marsiske,$^{2}$
M.~L.~Perl,$^{2}$ V.~Savinov,$^{2}$ D.~Ugolini,$^{2}$
X.~Zhou,$^{2}$
% ...
\end{center}
 
\small
\begin{center}
$^{1}${University of Rochester, Rochester, New York 14627}\\
$^{2}${Stanford Linear Accelerator Center, Stanford University, Stanford,
California 94309}\\
% ...
\end{center}

\setcounter{footnote}{0}
}
\newpage

%---------------------------------------------------------------------
%
\section{Introduction}
%
%---------------------------------------------------------------------

  The phenomenon of $CP$ violation, so far observed only in 
the neutral kaon system, can be accommodated  by a complex phase in the   
Cabibbo-Kobayashi-Maskawa (CKM) quark-mixing matrix~\cite{CKM}.
Whether this phase is the correct, or only, source of $CP$ violation
awaits experimental confirmation.
$B$ meson decays, in particular charmless $B$ meson decays,
will play an important role in verifying this picture.

\begin{figure}[hbp]
\centering
\leavevmode
\epsfxsize=6.5in
\epsffile{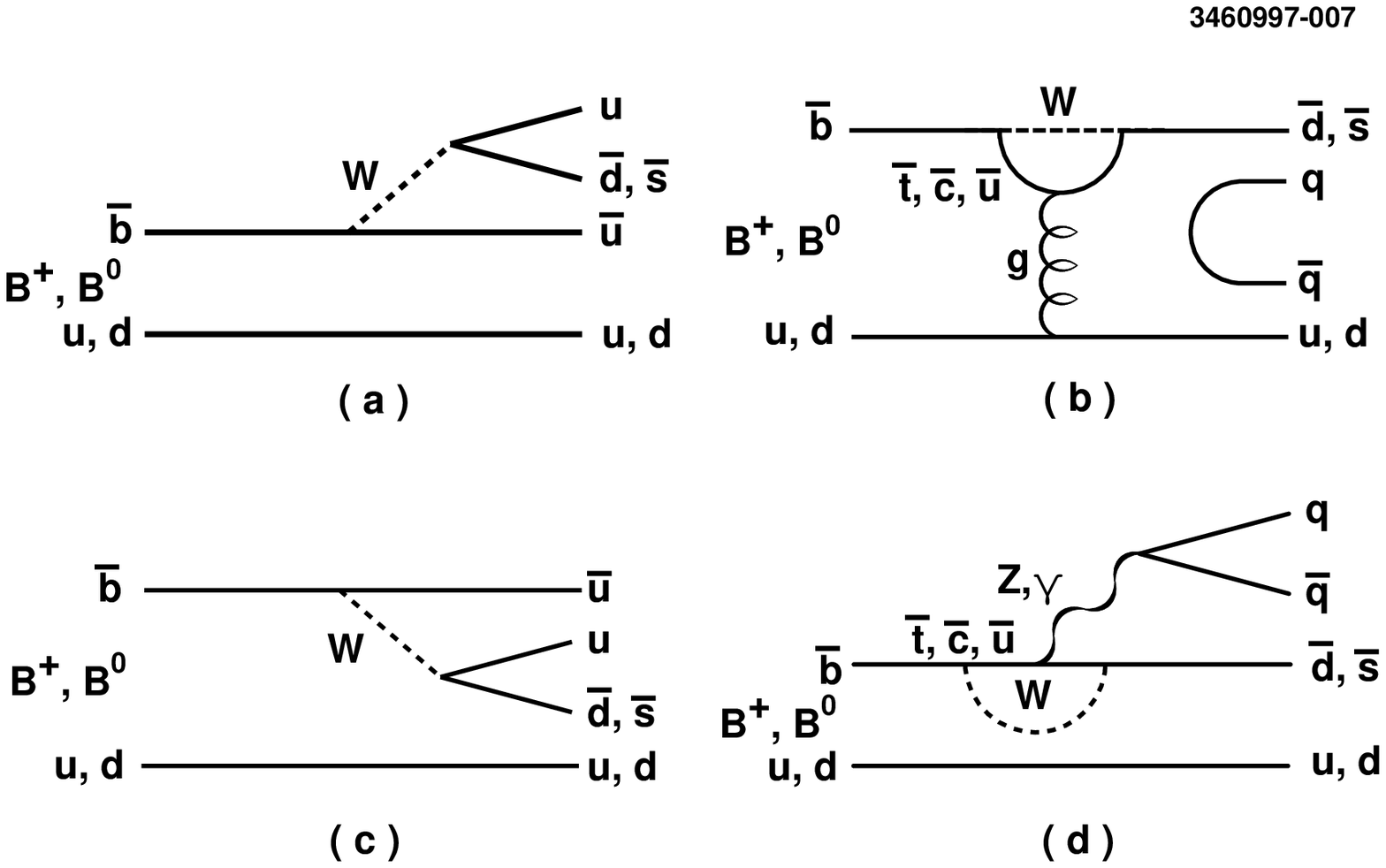}
\caption{The dominant decay processes are expected to be 
(a) external W-emission, (b) gluonic penguin, (c) internal W-emission,
(d) external electroweak penguin.}
\label{fig:feynman}
\end{figure}

The decay $B \rightarrow \pi^+\pi^-$, dominated by the $b \rightarrow
u$ tree diagram (Fig.~\ref{fig:feynman}(a)), can be used to measure $CP$ 
violation due to $B^0-\bar B^0$\ mixing at both asymmetric
$B$~factories and hadron colliders. 
However, theoretical uncertainties due to the presence of 
the $b\to dg$ penguin diagram (Fig.~\ref{fig:feynman}(b)) make it
difficult to extract the angle $\alpha$ of the unitarity triangle from 
$B \rightarrow \pi^+\pi^-$ alone. 
Additional measurements of 
$B^\pm \rightarrow \pi^\pm \pi^0$, $B,\bar{B} \rightarrow \pi^0\pi^0$, and the use
of isospin 
symmetry may resolve these uncertainties~\cite{isospin}.

 $B\to K\pi $\ decays are dominated by the
$b \rightarrow sg$ gluonic penguin diagram,
with additional contributions from 
$b \rightarrow u$ tree and color-allowed 
electroweak penguin (Fig.~\ref{fig:feynman}(d)) processes.
Interference between the penguin and tree amplitudes 
can lead to direct $CP$ violation, which
 would manifest itself as a rate asymmetry for
decays of $B$ and $\bar{B}$~mesons.
Several methods of measuring or constraining $\gamma$,  the phase of $V_{ub}$.
using only decay rates of $B\to K\pi,~\pi\pi$ processes were also 
proposed~\cite{triangles}~\cite{fleischer-mannel}~\cite{neubert-rosner}. 
This is particularly important, as $\gamma $\ is the
least known parameter of the 
unitarity triangle and is likely to remain
the most difficult to determine experimentally.
%The ratio
% $R={\cal B}(B\to K^{\pm}\pi^{\mp})/{\cal B}(B^{\pm}\to K^0\pi^{\pm}) $, 
%was recently suggested~\cite{fleischermannel} as a way to constrain $\gamma$. 
%However, final state interactions(FSI) in $B\to K\pi$ decays can
%significantly affect
%this method\cite{weyers}. Studies of B decays to the $KK$ final
%states can provide useful limits on the FSI effects\cite{gronaukk}.
 This paper presents the first observation of the decays $B\to
\pi^+\pi^-$ and $B\to K^0\pi^0$, improved measurements for 
$B\to K^{\pm}\pi^{\mp}$,
$B^{\pm}\to K^0\pi^{\pm}$, and $B^{\pm}\to K^{\pm}\pi^0$
 decays, and updated upper limits for $B$ decays
to $\pi^\pm \pi^0, ~K^+K^-$, and $K^0K^\pm$.
Average over charge conjugate decays is implied throughout this paper.

%---------------------------------------------------------------------
%
\section{Data Set, Detector, Event Selection}
%
%---------------------------------------------------------------------
The data used in this analysis was collected with the CLEO II
and CLEO II.V
detectors at the Cornell Electron Storage Ring (CESR).
It consists of $9.1~{\rm fb}^{-1}$ taken at the $\Upsilon$(4S)
(on-resonance) and $4.5~{\rm fb}^{-1}$ taken 
below $B\bar{B}$ threshold.  
The below-threshold sample
is used for continuum background studies.
The on-resonance sample contains 9.66~million $B\bar{B}$ pairs.
This is a $67\% $ increase in the number of $B\bar{B}$ pairs
over the previously presented analysis~\cite{ichep98}.

 CLEO II and CLEO II.V are 
general purpose solenoidal magnet detectors, described in 
detail elsewhere~\cite{detector}.
 In CLEO II,
 the momenta of charged particles are measured in a
tracking system consisting of a 6-layer straw
tube chamber, a 10-layer precision
drift chamber, and a 51-layer main drift chamber, all operating
inside a 1.5 T superconducting solenoid.  The main drift chamber
also provides a measurement of the specific ionization loss, $dE/dx$,
used for particle identification.
 For CLEO II.V the 6-layer straw tube chamber was replaced by a 3-layer
double-sided silicon vertex detector, and the gas in the main 
drift chamber was changed from an argon-ethane to a helium-propane
mixture. 
%As a result, 
%we have improved our $dE/dx$ and momentum resolution
%for charged kaons
%and pions in $B\to K^\pm\pi^\mp$ by 18\% in CLEO II.5 over CLEO II.
Photons are detected
using 7800-crystal CsI(Tl) electromagnetic calorimeter. Muons are 
identified using proportional counters placed at various depths in the
steel return yoke of the magnet.

Charged tracks are
required to pass track quality cuts based on the average hit residual
and the impact parameters in both the $r-\phi$ and $r-z$ planes.
%Pairs of tracks with vertices 
%displaced by at least $3$~mm from the primary
%interaction point are taken as $K_S^0$ candidates. 
Candidate $K^0_S$ are selected from pairs of tracks forming well-measured 
displaced vertices.
%a flight distance significance with respect to the beam spot of at least
%$3\sigma$ for CLEO II and $5.5\sigma$ for CLEO II.5. 
Furthermore, we require
the $K^0_S$ momentum vector to point back to the beam spot and 
the $\pi^+\pi^-$ invariant mass to be within $10$~MeV, $\sim 2.5$ standard
deviations ($\sigma$), 
of the $K^0_S$\ mass.
Isolated showers with energies greater than
$40$~MeV in the central region of the CsI calorimeter
and greater than $50$~MeV elsewhere, are defined to be photons.
To reduce combinatoric backgrounds we
require the
lateral shapes of the showers to be consistent with those from photons. 
To suppress further low energy showers from charged particle interactions
in the calorimeter we apply a shower energy dependent isolation cut.
Pairs of photons with an invariant mass within $25$~MeV ($\sim 2.5\sigma$)
of the nominal $\pi^0$ mass
are kinematically fitted with the mass constrained to the
$\pi^0$ mass.  

Charged particles ($h^\pm$) are identified as kaons or pions using $dE/dx$.
Electrons are rejected
based on $dE/dx$ and the ratio of the track momentum to the
 associated shower energy in the CsI calorimeter.
We reject muons
by requiring that the tracks do not penetrate the steel absorber to a
depth greater than seven nuclear interaction lengths.
We have studied the $dE/dx$\ separation between 
kaons and pions for momenta $p \sim 2.6$~GeV$/c$\ in data using
$D^{*+}$-tagged 
$D^0\rightarrow K^- \pi^+$ decays; we find a separation of 
$(1.7\pm 0.1)~\sigma$
for CLEO II and $(2.0\pm 0.1)~\sigma$ for CLEO II.V.

%---------------------------------------------------------------------
%
\section{Analysis}
%
%---------------------------------------------------------------------
We calculate a beam-constrained $B$ mass 
$M = \sqrt{E_{\rm b}^2 - p_B^2}$, where $p_B$ is the $B$
candidate momentum and $E_{\rm b}$ is the beam energy.
The resolution in $M$\ is dominated by the beam energy spread and
ranges from 2.5 to 3.0~${\rm MeV}/{\it c}^2$, 
where the 
larger resolution corresponds to decay modes with a $\pi^0$.
We define $\Delta E = E_1 + E_2 - E_{\rm b}$, where $E_1$ and $E_2$
are the energies of the daughters of the $B$ meson candidate.
The resolution on $\Delta E$ is mode dependent.
For final states without $\pi^0$'s, the $\Delta E$ resolution for CLEO II(II.V)
is $\pm 25(20)$~MeV. In the 
$B^\pm\to h^\pm \pi^0$ analysis the $\Delta E$ 
resolution is
worse by approximately a factor of two and becomes asymmetric because of
energy loss out of the back of the CsI crystals. 
%In $B\rightarrow K^+\pi^0/\pi^+\pi^0$ we parameterize 
%the $\Delta E$ resolution
%by a sum of two bifurcated Gaussians of widths $+35/-41(+125/-60)$MeV,
%with 68\% of the total area in the narrower of the two Gaussians.
%The resolution is asymmetric because of energy loss out of the
%back of the CsI crystals. 
The energy constraint also helps to distinguish between modes of
the same topology.  
For example, $\Delta E$
for  $B \rightarrow K^+ \pi^-$,  calculated assuming 
$B \rightarrow \pi^+\pi^-$,
has a distribution that is centered at $-42$~MeV, giving a separation
of $1.7(2.1)\sigma$ between $B \rightarrow K^+ \pi^-$ and
$B \rightarrow \pi^+\pi^-$ for CLEO II(II.V).
We accept events with $M$\ within $5.2-5.3$~$\rm {GeV/c^2}$\ and 
$|\Delta E|<200(300)$~MeV
for decay modes without (with) a $\pi^0$\ in the final state. This fiducial
region 
includes the signal region and a sideband for background determination.

We have studied backgrounds from $b\to c$\ decays and other $b\to u$\ and
$b\to s$\ decays and find that all are negligible for the analyses presented
here. The main background arises from $e^+e^-\to q\bar q$\ (where $q=u,d,s,c$).
Such events typically exhibit a two-jet structure and can produce high 
momentum back-to-back tracks in the fiducial region.
To reduce contamination from these events, we calculate the angle $\theta_S$
between the sphericity axis\cite{shape} of the candidate tracks and showers
and the
sphericity axis of the rest of the event. The distribution of $\cos\theta_S$\ 
is strongly peaked at $\pm 1$ for $q\bar q$\ events and is nearly flat 
for $B\bar B$\ 
events. We require $|\cos\theta_S|<0.8$\ which eliminates  $83\%$\ 
of the background. 
Using a detailed GEANT-based Monte-Carlo simulation~\cite{geant}
we determine overall detection efficiencies (${\cal E}$) of $11-45\%$,
as listed in
Table~\ref{tab:results}. Efficiencies include the branching fractions for
$K^0\to K^0_S\to \pi^+\pi^-$\ and $\pi^0\to \gamma\gamma$ where applicable.

%We estimate a systematic error on the efficiency using
%independent data samples.

Additional discrimination between signal and $q\bar q$\ background is provided
by a Fisher discriminant technique as described in detail in 
Ref.~\cite{bigrare}.
The Fisher discriminant is a linear combination,
${\cal F}\equiv \sum_{i=1}^{N}\alpha_i y_i$,\ where the coefficients 
$\alpha_i$ are chosen to maximize the separation between the signal
and background Monte Carlo samples. 
The 11 inputs, $y_i$, are the cosine of the angle 
between the candidate sphericity axis and beam axis, the
ratio of Fox-Wolfram moments $H_2/H_0$~\cite{fox}, and nine variables that 
measure the scalar sum of the momenta of tracks and showers from the rest of 
the event in 
nine angular bins, each of $10^\circ$, centered about the candidate's
sphericity axis.

We perform unbinned
maximum-likelihood (ML) fits using $\Delta E$, $M$, ${\cal F}$,
% $|\cos\theta_B|$ 
the angle between the $B$ meson momentum and beam axis, 
and $dE/dx$ (where applicable) as input information for each candidate
event to determine the signal yields. 
Four different fits are performed, one for each topology 
($h^+h^-$, $h^\pm \pi^0$, $h^\pm K^0_S $, and $K^0_S\pi^0$, 
$h^\pm$\ referring to a charged kaon or pion).
In each of these fits, the likelihood of the  event
is parameterized by the sum of probabilities for
all relevant signal and background hypotheses,
with relative weights determined by maximizing the likelihood function 
($\cal L$).
The probability of a particular hypothesis is calculated as a product of the
probability density functions (PDFs) for each of the input variables.
Further details about the likelihood fit can be found in Ref.~\cite{bigrare}.
The parameters for the PDFs 
are determined from independent data and high-statistics Monte Carlo
samples. We estimate a systematic error on the fitted yield by varying the
PDFs used in the fit within their uncertainties.
These uncertainties are dominated by the limited statistics in the
independent data samples we used to determine the PDFs. 
The systematic errors on the measured branching fractions are obtained
by adding this fit systematic in quadrature with the systematic error
on the efficiency.
%\begin{figure}[htbp]
%\centering
%\leavevmode
%\epsfxsize=3.5in
%\epsffile{/cdat/lnsct2/disk1/fkw/rareb/maxl/data/pdfs/hpzfdsystlog.ps}
%\caption{Variations in Fisher discriminant background  PDF
%used to estimate systematic error in $h\pi^0$ analysis.}
%\label{fig:fdtail}
%\end{figure}

%To evaluate how systematic uncertainties in the PDFs
%affect the statistical
%significance of our signals, we repeated the fits for the
%$h^+h^-,~h^+\pi^0$, and $h^+K^0_S$ modes with all PDFs 
%changed within their uncertainty to maximally reduce the overall
%signal yield. Under these conditions, the significance of the $K^+\pi^-, 
%~K^+\pi^0$, and $K^0_S\pi^+$ signals becomes $6.7,~4.7$, and
%$4.2~\sigma$ respectively.

%---------------------------------------------------------------------
%
\section{Results}
%
%---------------------------------------------------------------------

We summarize all branching fractions and upper limits in Table
\ref{tab:results}.  We find statistically significant signals for the
decays $B \to K^\pm \pi^\mp$, $B\to\pi^+\pi^-$, $B^\pm \to K^\pm \pi^0$,
$B^\pm \to K^0_S \pi^\pm$, and $B \to K^0_S \pi^0$.

\begin{figure}[htbp]
\centering
\leavevmode
\epsfxsize=6.0in
\epsffile{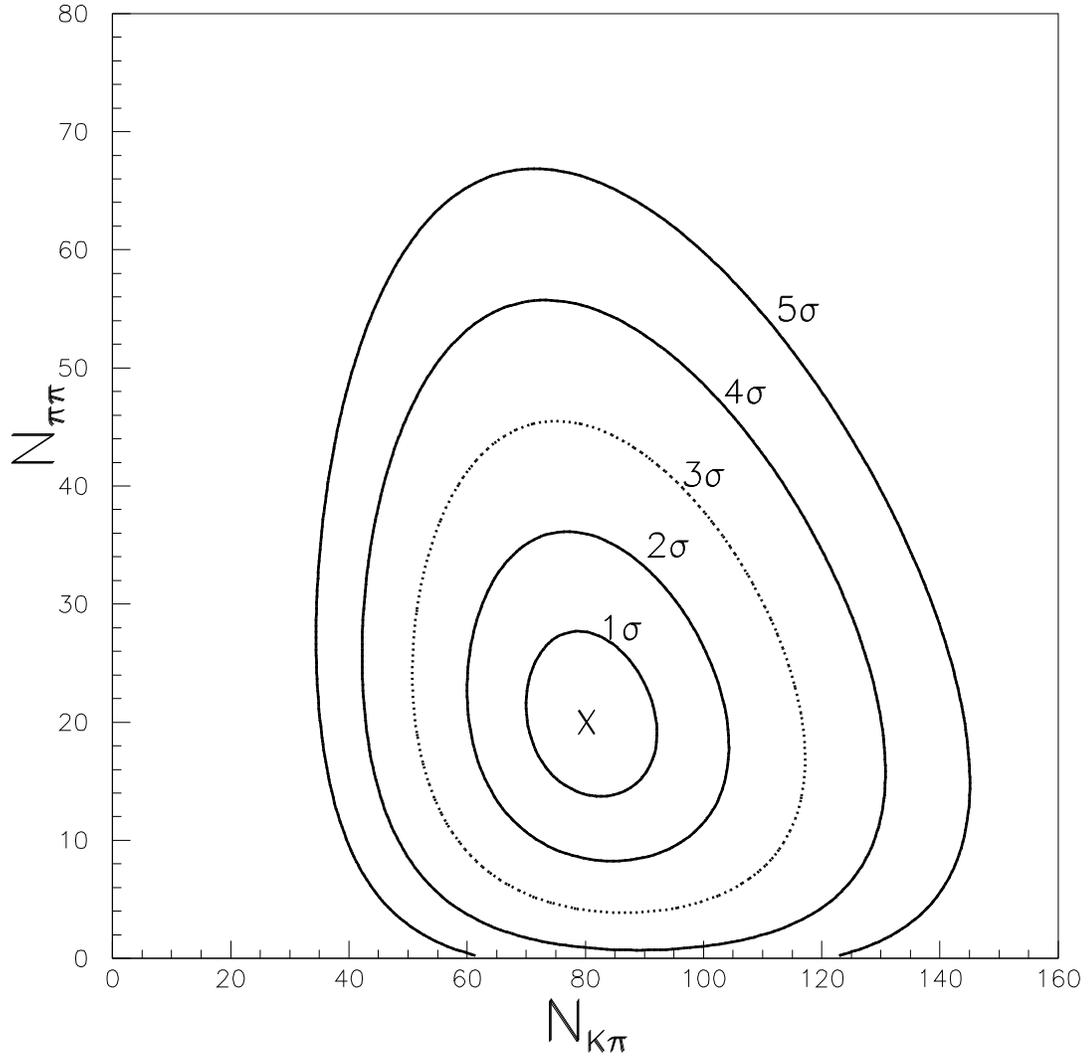}
\caption{Contours of the $-2\ln{\cal L}$ for the ML fit to 
$N_{K^{\pm}\pi^{\mp}}$ and $N_{\pi^+\pi^-}$, the $K^\pm\pi^\mp$
and $\pi^+\pi^-$ yields respectively. }
\label{fig:contourkpi}
\end{figure}

\begin{figure}[htbp]
\centering
\leavevmode
\epsfxsize=6.5in
\epsffile{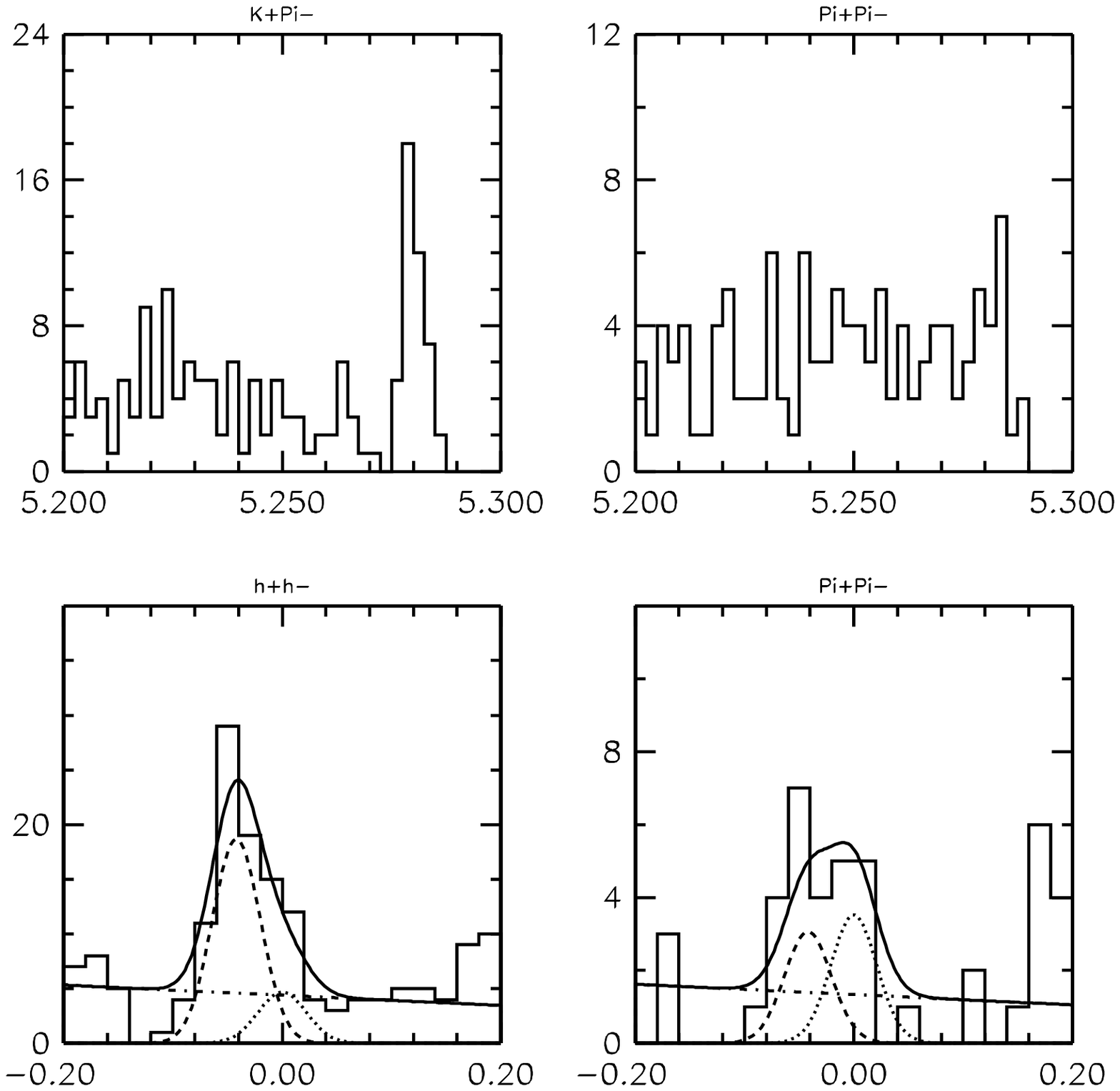}
\caption{Projections of $K\pi$ and $\pi\pi$ events
onto $M$ and $\Delta E$ with cuts.
Upper left: $M$ distribution of $K\pi$-like events;
upper right: $M$ distribution of $\pi\pi$-like events.
Lower left: $\Delta E$ distribution of events prior
to $\pi\pi$ vs $K\pi$ vs $KK$ selection according to dE/dx;
Lower right: $\Delta E$ distribution of events that are
more likely to be $\pi\pi$ than $K\pi$ or $KK$ based on dE/dx.
Overlays in the lower plots are the results of the likelihood
fit scaled by the efficiency of the cuts used to project
into these plots. Solid line: total fit; dashed: $K\pi$;
dotted: $\pi\pi$; dot-dash: continuum background.}
\label{fig:kpi-pipi-projections}
\end{figure}

\begin{table}
\begin{center}
\caption{Experimental results. 
Branching fractions
(${\cal B}$) and 90\% C.L. upper limits are given in units
of $10^{-6}$.
The errors on ${\cal B}$ are statistical and 
systematics respectively. Reconstruction efficiency ${\cal E}$
includes branching fractions of $K^0 \to K^0_S \to \pi^+\pi^-$ and 
$\pi^0\to \gamma\gamma$. We quote the central value branching fraction
in $\pi^\pm\pi^0$ for convenience only. The
statistical significance of the excess above background in this 
final state is insufficient for a first observation
of this decay mode.}
% should we quote raw yields instead? B_fit looks ugly.
\begin {tabular}{l c c c c }
Mode& ${\cal E}(\%)$ & ${\cal B}_{fit}$($10^{-6}$) & Signif.(std.dev.) 
&${\cal B}$($10^{-6}$) \\
\hline
$\pi^+\pi^-$ &
$45$
& $4.7^{+1.8}_{-1.5}$  & 4.2  
& $4.7^{+1.8}_{-1.5}\pm 0.6$       \\
$\pi^+\pi^0$ & 
$41$
& $5.4^{+2.1}_{-2.0}\pm 1.5$ & 3.2 
&$<12$         \\
\hline
$K^+\pi^-$   &
$45$
& $18.8^{+2.8}_{-2.6}$ & 11.7
& $18.8^{+2.8}_{-2.6} \pm 1.3$        \\
$K^+\pi^0$   &
$38$
& $12.1^{+3.0}_{-2.8}$  & 6.1
& $12.1^{+3.0+2.1}_{-2.8-1.4}$         \\
$K^0\pi^+$   
& $14$
&$18.2^{+4.6}_{-4.0}$ & 7.6
&$18.2^{+4.6}_{-4.0} \pm 1.6$               \\
$K^0\pi^0$   
& $11$
&$14.8^{+5.9}_{-5.1}$ & 4.7
&$14.8^{+5.9+2.4}_{-5.1-3.3}$               \\
\hline
$K^+K^-$     
& $45$
&  & 0.
& $<2.0$  \\
$K^+\bar{K}^0$   
& $14$
% HAVE TO DECIDE WHAT TO QUOTE FOR CENTRAL VALUE AND ITS ERRORS
& $$  & 1.1 
& $<5.1$    \\
\end {tabular}
\label{tab:results}
\end{center}
\end {table}
%-------------------------------------------------------------------
% pipi
%-------------------------------------------------------------------
Fig. \ref{fig:contourkpi} shows the results of the likelihood fit for
$B\to \pi^\pm\pi^\mp$ and $B\to K^\pm\pi^\mp$.  The curves represent
the $n\sigma$ contours, which correspond to the increase in
$-2\ln{\cal L}$ by $n^2$. The dashed curve marks the $3\sigma$ contour;
systematic uncertainties are not included in any contour plots.  The
statistical significance of a given signal yield is determined by
repeating the fit with the signal yield fixed to be zero and recording
the change in $-2\ln{\cal L}$.  Fig. \ref{fig:kpi-pipi-projections}
shows distributions in $M$ and $\Delta E$ for events after cuts on the
Fisher discriminant and whichever of $M$ and $\Delta E$ is not being
plotted, plus an exclusive classification into $K\pi$-like and $\pi\pi$-like
candidates
based on the most probable assignment with $dE/dx$ information. The
likelihood fit, suitably scaled to account for the efficiencies of the
additional cuts, 
is overlaid in the $\Delta E$ distributions to illustrate the separation
between $K\pi$ and $\pi\pi$ events.

%/cdat/df/pg/ps/lp99/contkpi_prl.ps

%/cdat/df/pg/ps/lp99/pipi_lp99.ps

\begin{figure}[htbp]
\centering
\leavevmode
\epsfxsize=6.0in
\epsffile{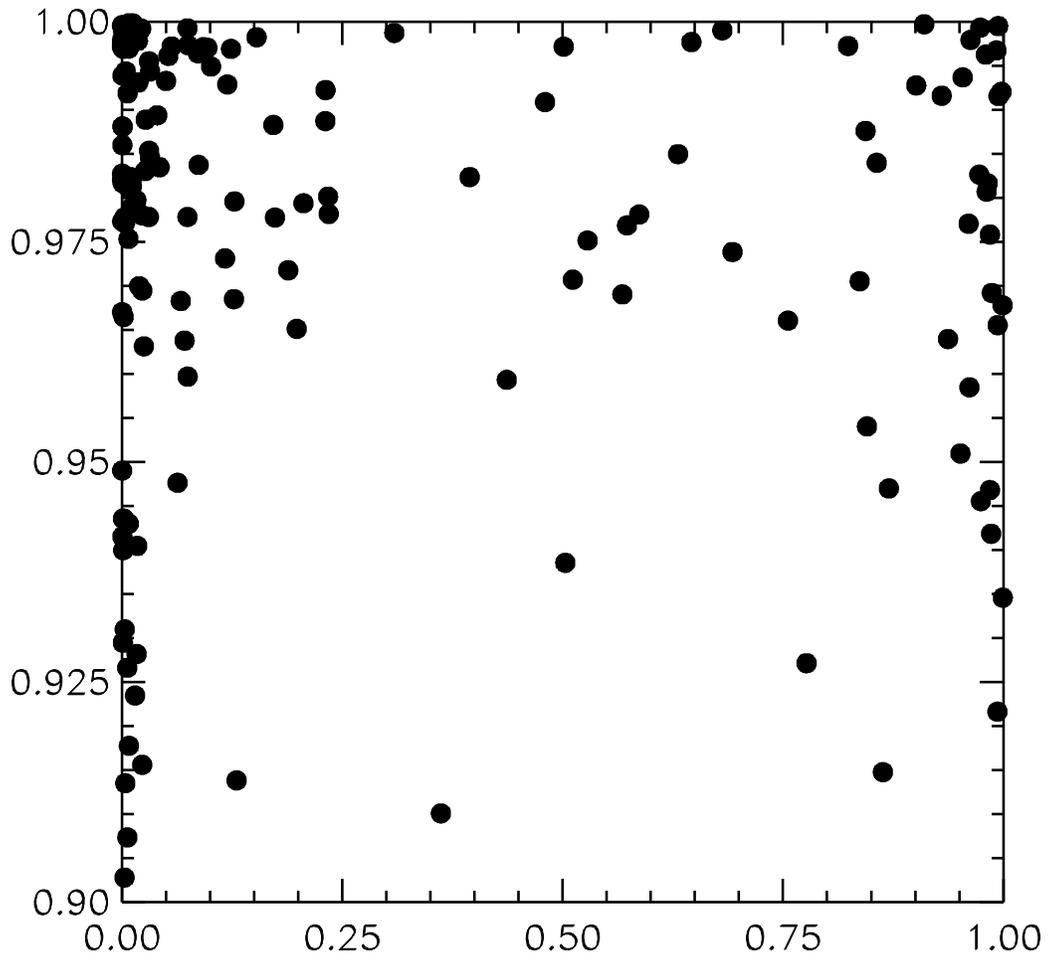}
\caption{The horizontal axis
shows $P^s_{\pi\pi}/(P^s_{\pi\pi}+P^s_{K\pi})$ while the 
vertical axis depicts
$(P^s_{\pi\pi}+P^s_{K\pi})/
(P^s_{\pi\pi}+P^s_{K\pi}+P^c_{\pi\pi}+P^c_{K\pi}+P^c_{KK})$.
Superscript $s$ and $c$ stand for signal and continuum background, 
respectively.
Signal events cluster near the top of the figure, and
separate into $K\pi$-like events on the left and $\pi\pi$-like
events on the right.}
\label{fig:likeratio}
\end{figure}

We also compute from the PDFs the event-by-event probability to
be signal or continuum background, and also the probability to
be $K\pi$-like or $\pi\pi$-like. From these we form likelihood
ratios, 
${\cal R}_{sig} = 
(P^s_{\pi\pi}+P^s_{K\pi})/
(P^s_{\pi\pi}+P^s_{K\pi}+P^c_{\pi\pi}+P^c_{K\pi}+P^c_{KK})$
and
${\cal R}_{\pi} = P^s_{\pi\pi}/(P^s_{\pi\pi}+P^s_{K\pi})$.
Superscript $s$ and $c$ stand for signal and continuum background 
respectively.
Fig. \ref{fig:likeratio} illustrates the distribution of events
in ${\cal R}_{sig}$ (vertical axis) and ${\cal R}_\pi$ (horizontal
axis).  Signal events cluster near the top of the figure, and
separate into $K\pi$-like events on the left and $\pi\pi$-like
events on the right.

We find no evidence for the decay $B\to K^\pm K^\mp$ and set an
upper limit accordingly, as shown in Table \ref{tab:results}.

%/cdat/df/pg/ps/lp99/kpi_pipi_ritchie.ps

%-------------------------------------------------------------------
% Kspi0
%-------------------------------------------------------------------
The results of the $K_S\pi^0$ fit are
shown in Fig. \ref{fig:mass_kspi0}.  The signal yield of 
$15.5^{+5.9}_{-5.0}$ events is 4.7$\sigma$ significant and
robust under variations of cuts and PDF parameter variations.

%/cdat/df/jima/rareps/2lnL.eps
\newcommand{\fisher}    {\mbox{${\cal F}$}}
\begin{figure}[p]
\begin{center}
\leavevmode
\epsfxsize=2.7in
\epsffile{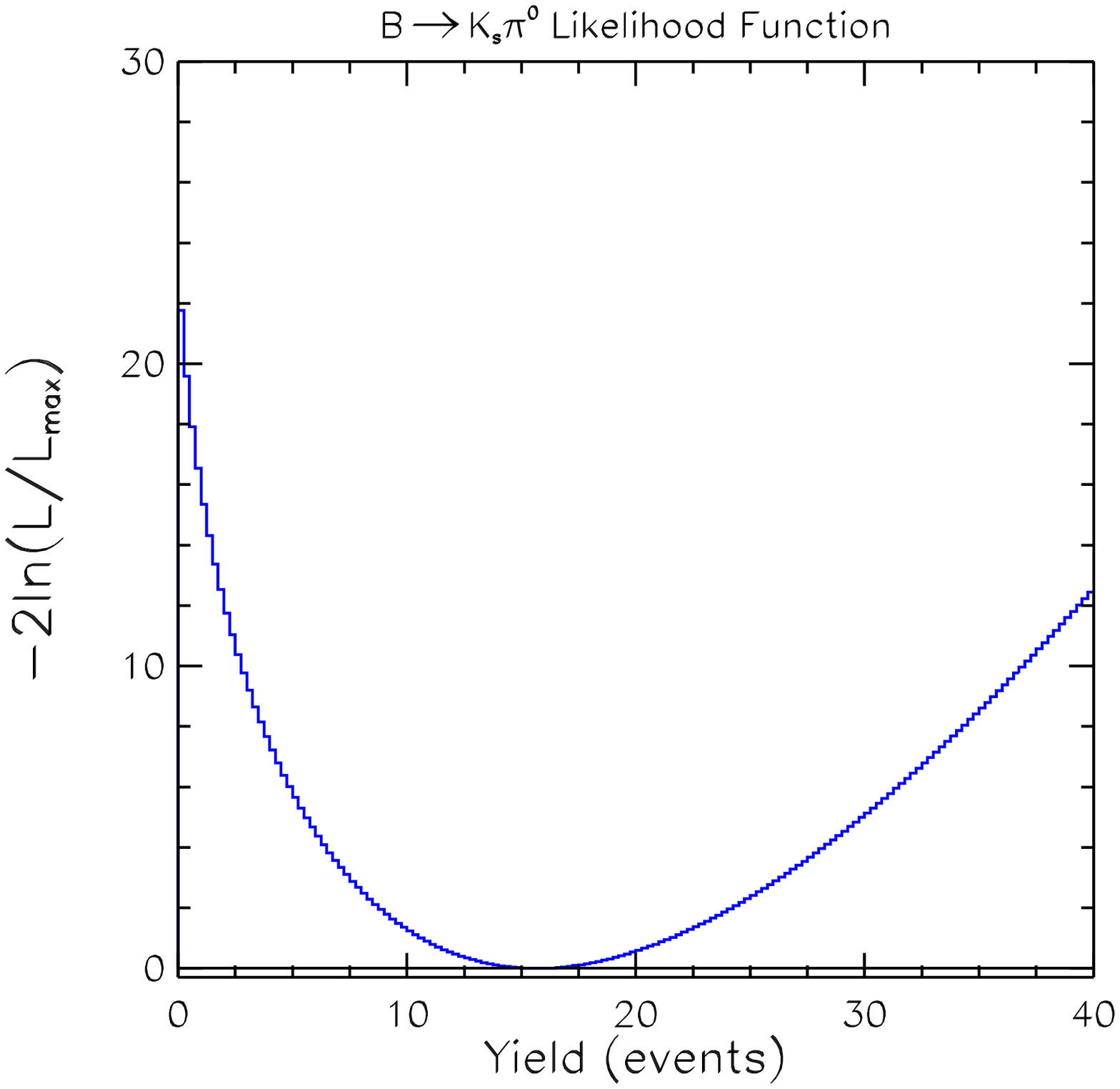}
\epsfxsize=2.7in
\epsffile{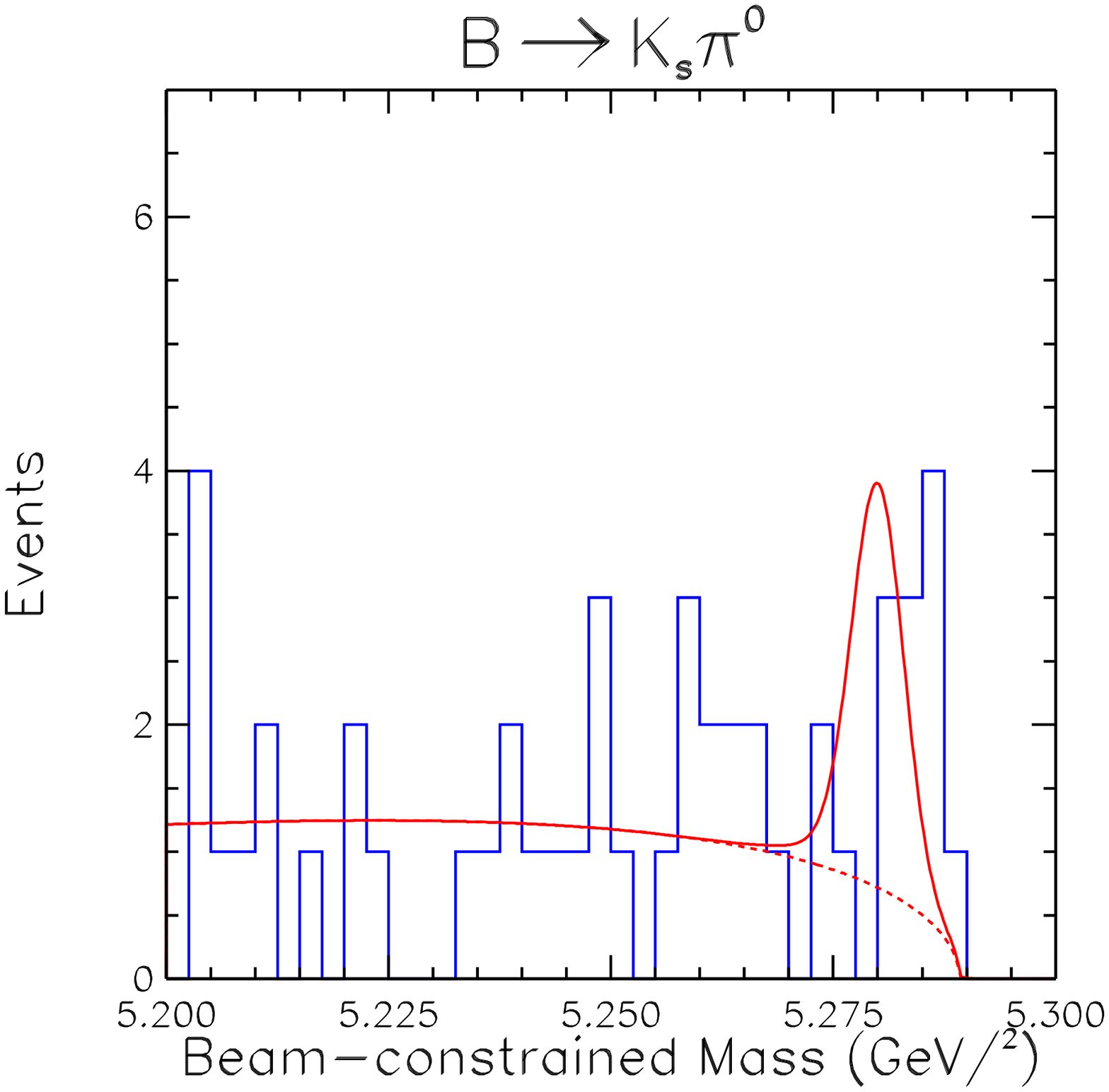}\\
\epsfxsize=2.7in
\epsffile{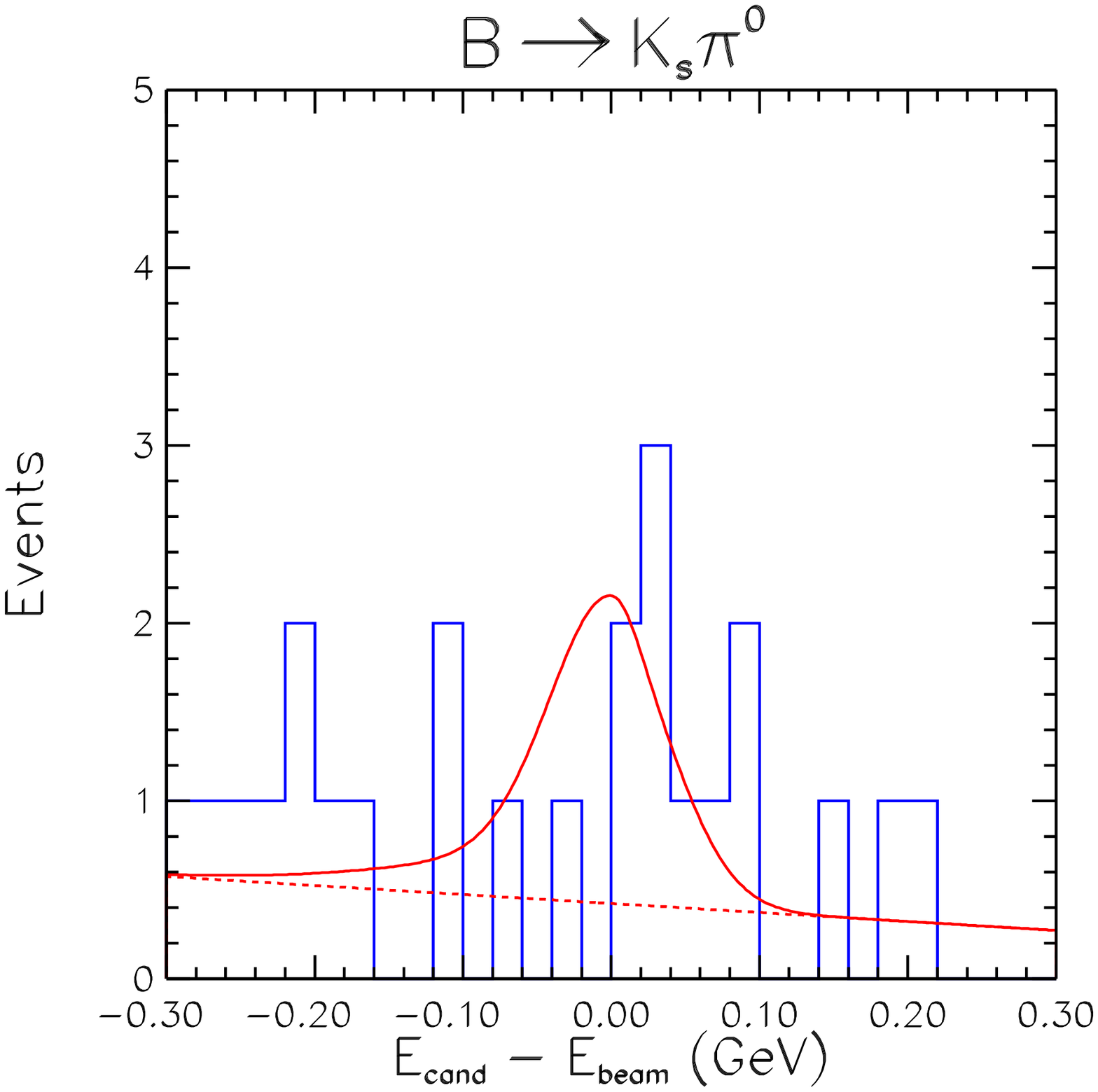}
\epsfxsize=2.7in
\epsffile{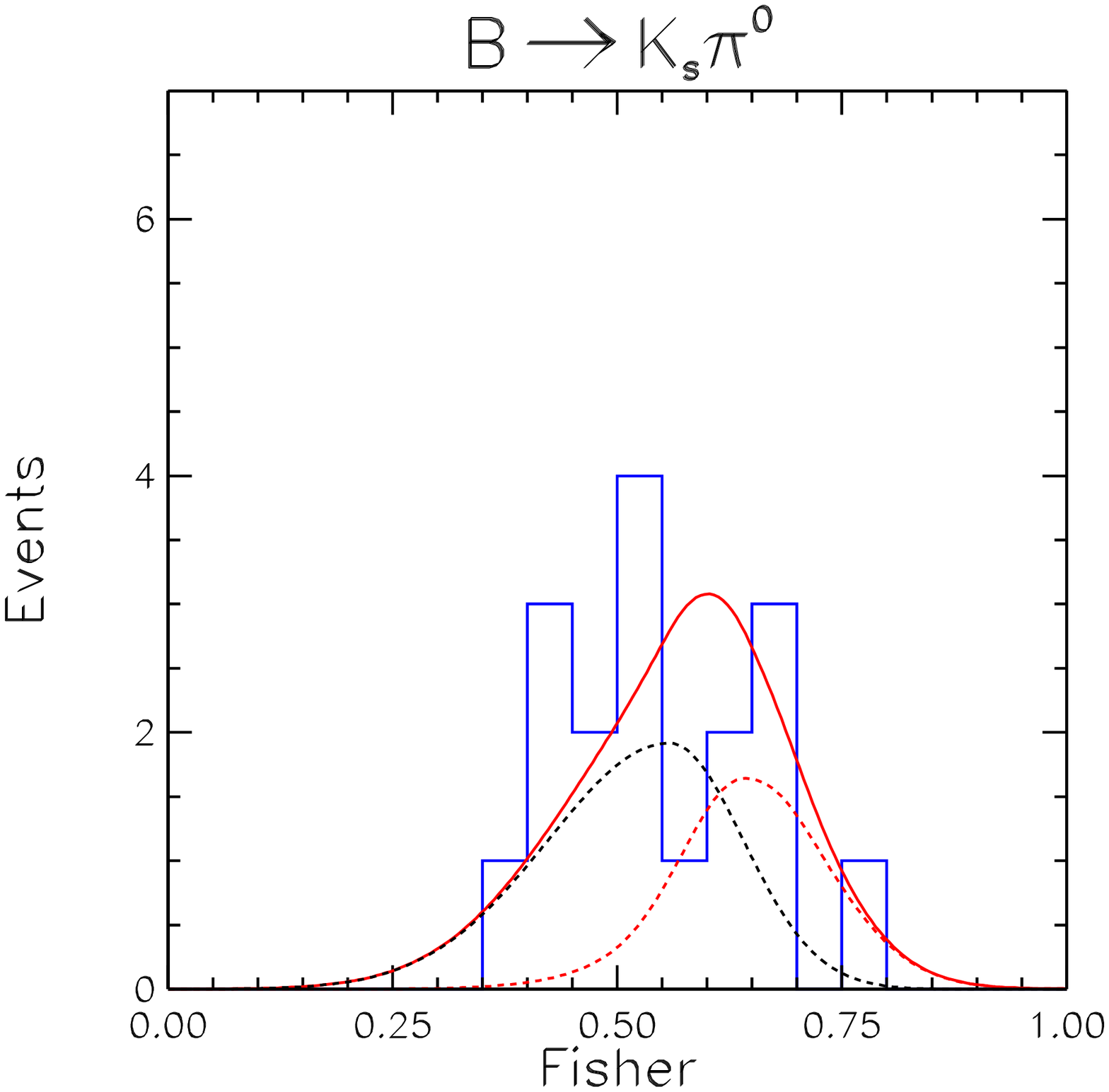}
\caption{$B\to K_S\pi^0$: 
Upper left: Likelihood function versus $K_S\pi^0$ yield;
Upper right: projection onto $M$ axis (efficiency = 0.63);
Lower left: projection onto $\Delta E$ axis (efficiency = 0.63);
Lower right: projection onto \fisher\ axis (efficiency = 0.69).
Overlaid curves are the results of the likelihood fit, scaled
according to the efficiency of the projection cuts.  Solid: total
fit; dotted: continuum background; dashed: signal.}
\label{fig:mass_kspi0}
\end{center}
\end{figure}

\begin{figure}[htbp]
\centering
\leavevmode
\epsfxsize=6.0in
\epsffile{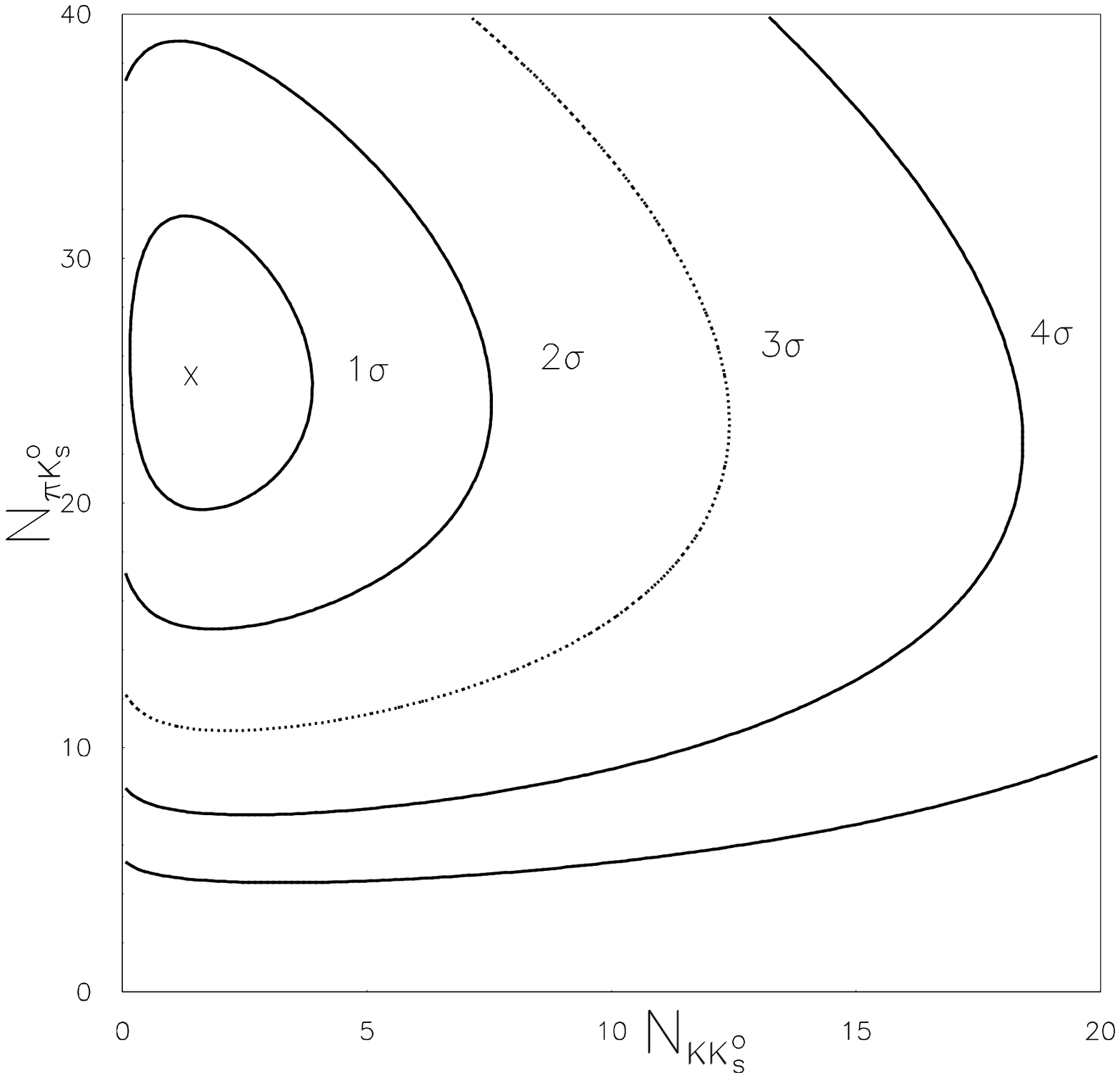}
\caption{Contours of the $-2\ln{\cal L}$ for the ML fit to 
$N_{K^0_S\pi^{\pm}}$ and $N_{K^0_SK^{\pm}}$. }
\label{fig:contourtks}
\end{figure}

\begin{figure}[htbp]
\centering
\leavevmode
\epsfxsize=6.0in
\epsffile{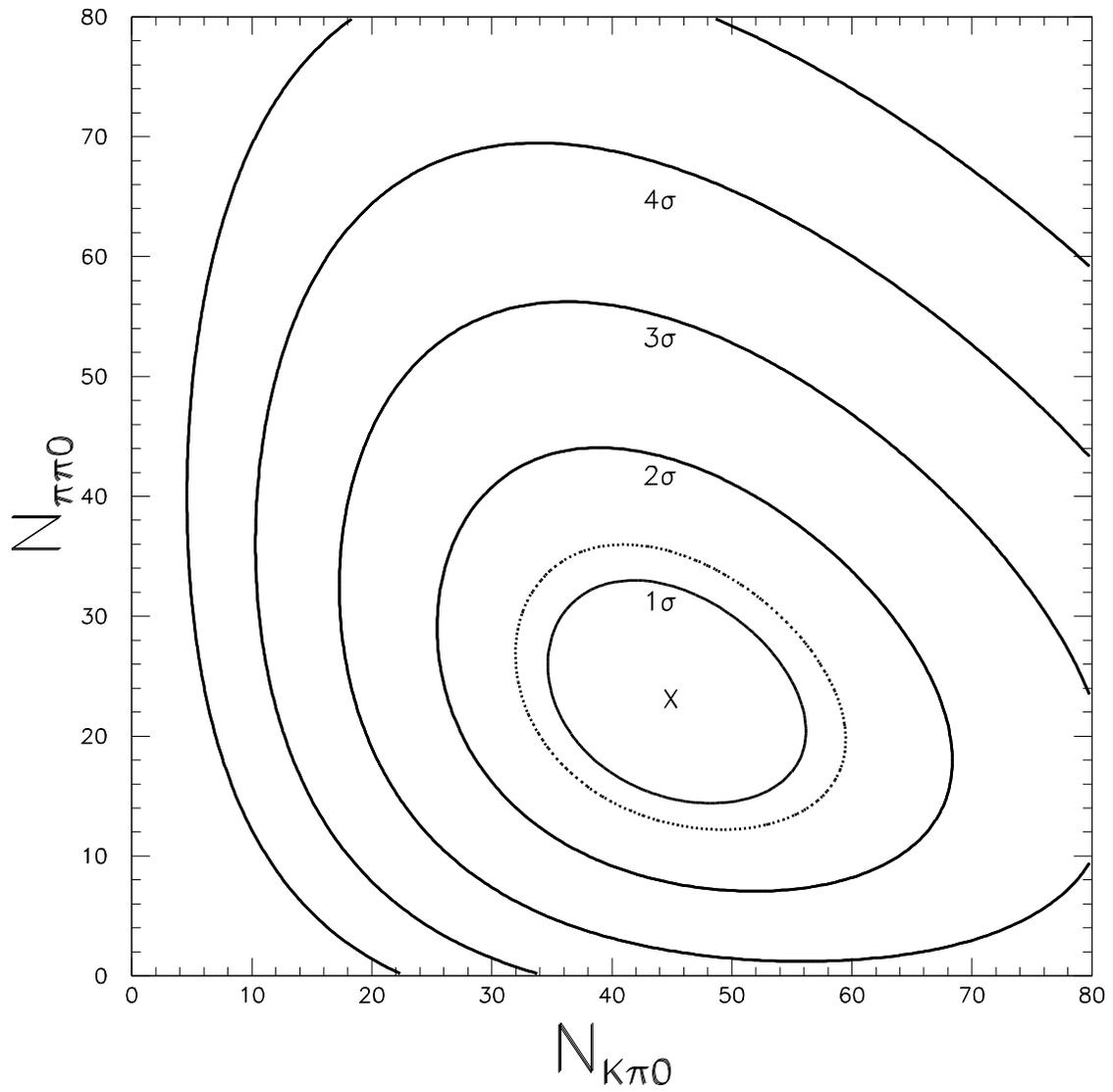}
\caption{Contours of the $-2\ln{\cal L}$ for the ML fit to 
$N_{K^{\pm}\pi^{0}}$ and $N_{\pi^{\pm}\pi^{0}}$. }
\label{fig:contourkpi0}
\end{figure}

\begin{figure}[htbp]
\centering
\leavevmode
\epsfxsize=6.5in
\epsffile{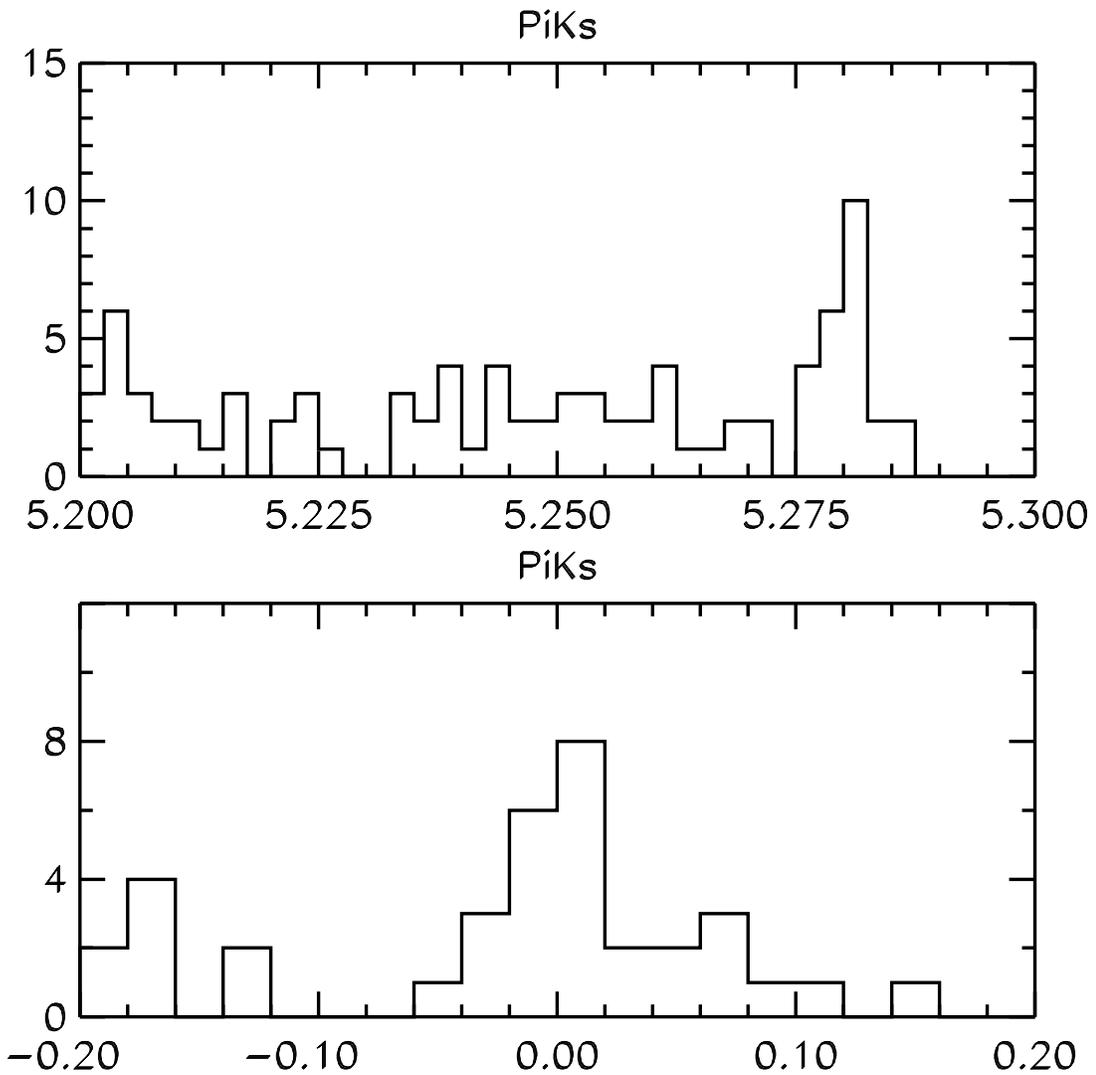}
\caption{$B\to K_S\pi^\pm$. Projections onto $M$ (top) and 
$\Delta E$ (bottom) in GeV.
Cuts are made on the Fisher discriminant
and whichever of $M$ and $\Delta E$ is not
being plotted.}
\label{fig:kspi_mb_de}
\end{figure}

\begin{figure}[htbp]
\centering
\leavevmode
\epsfxsize=3.0in
\epsffile{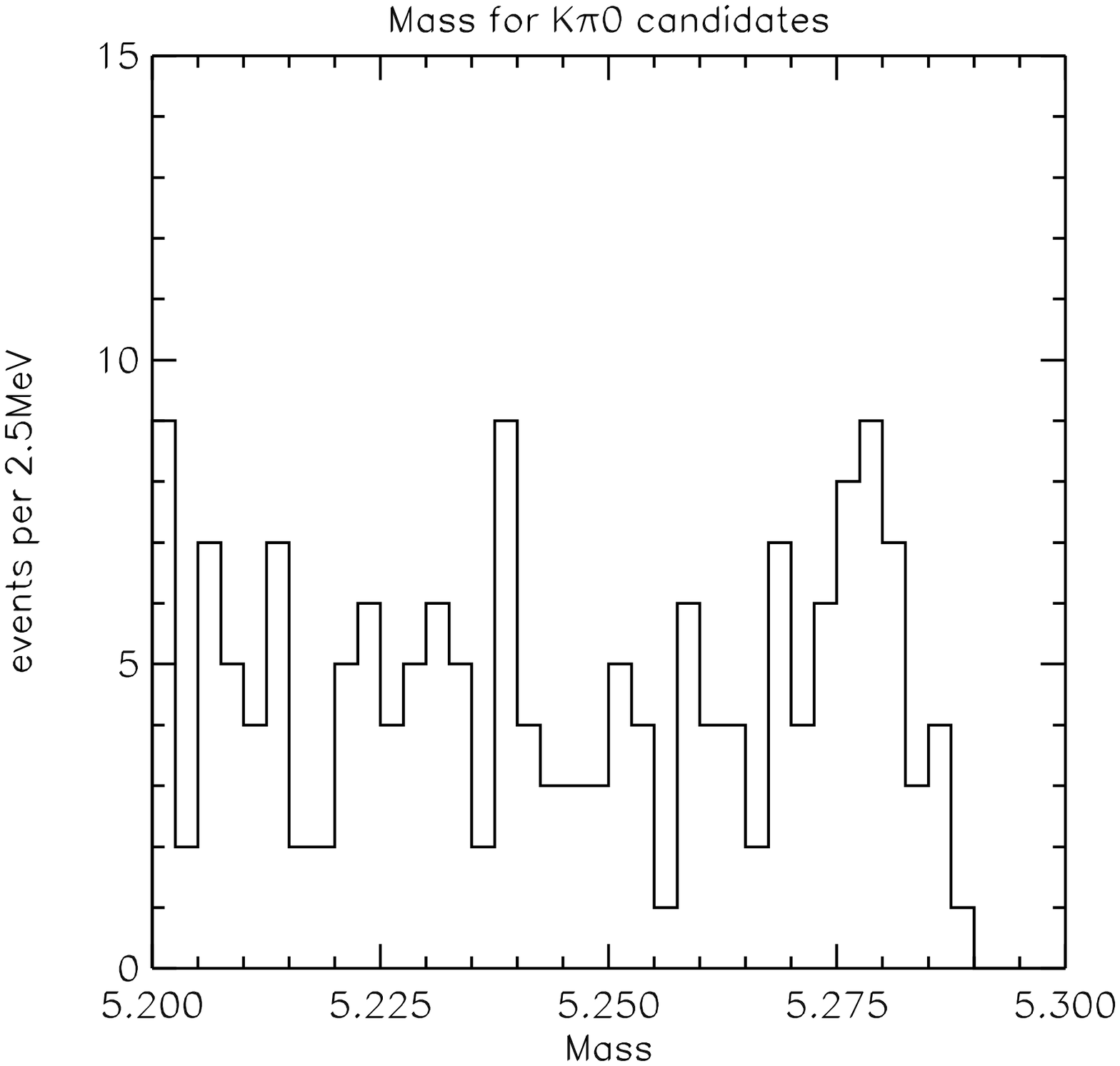}
\epsfxsize=3.0in
\epsffile{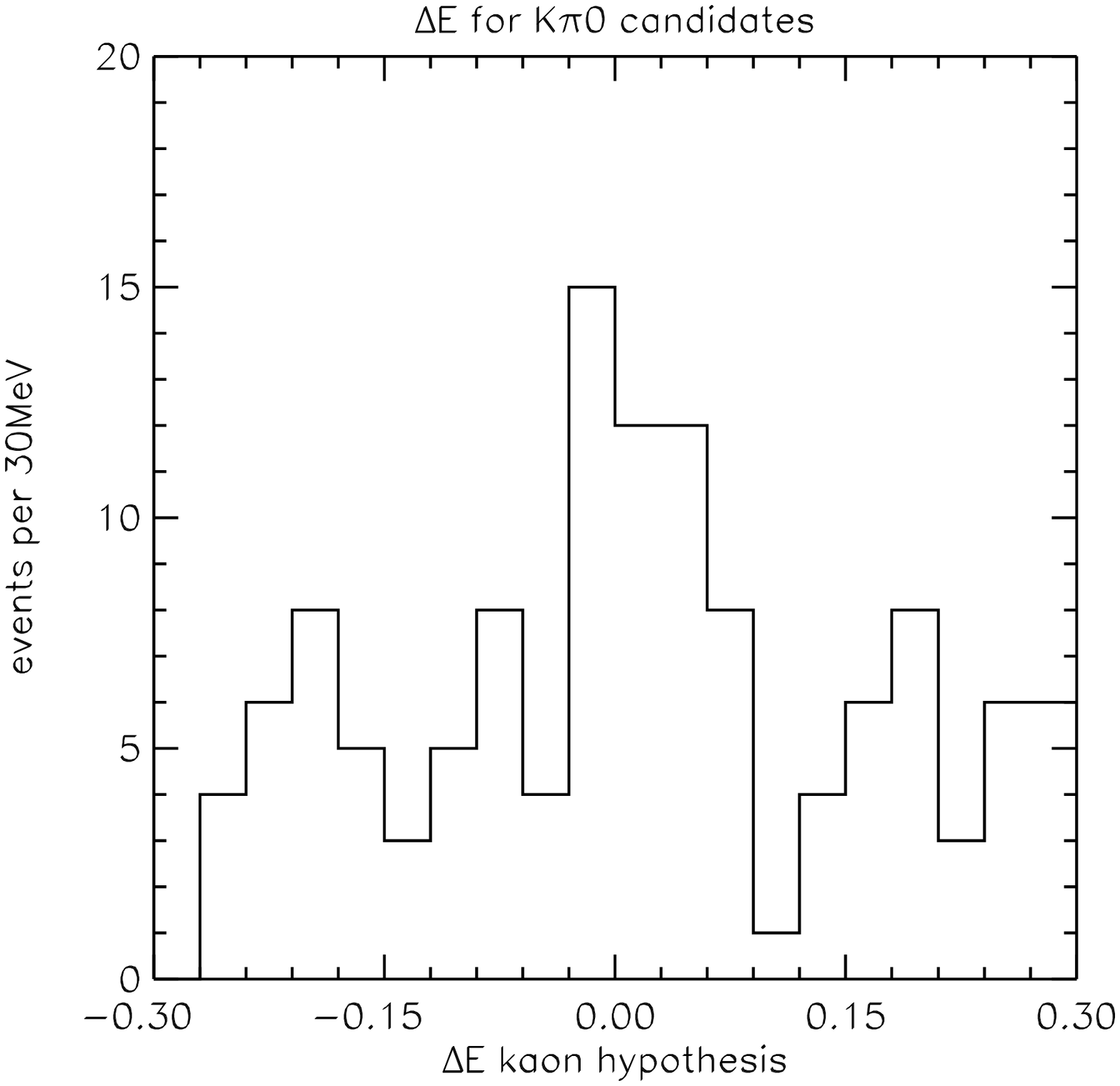}
\caption{$B\to K^\pm\pi^0$. Projections onto $M$ (left) and 
$\Delta E$ (right).
Cuts are made on the Fisher discriminant
and whichever of $M$ and $\Delta E$ is not
being plotted.}
\label{fig:kpi0_mb_de}
\end{figure}

%-------------------------------------------------------------------
% Everything else.
%-------------------------------------------------------------------
Figures \ref{fig:contourtks} and \ref{fig:contourkpi0} illustrate
contour plots of $-2\ln{\cal L}$ for the ML fit to $K^0_sh^\pm$ and
$h^\pm \pi^0$. The branching ratios and limits associated
with these four fits are given in Table \ref{tab:results}.
We also show projections onto $M$ and $\Delta E$ for these modes
in Figures \ref{fig:kspi_mb_de} and \ref{fig:kpi0_mb_de}.

%  To further illustrate the fits,
%Figures~\ref{fig:mass_hh}, and \ref{fig:mass_hks}
%show $\Delta E$ projections for events
%in a signal region defined by
%$|\Delta E| < 2\sigma_{\Delta E}$ ( $|M-5.28| < 2\sigma_{M}$) for
%the final states $K^\pm \pi^\mp,~K^\pm \pi^0$, and $K^0_S \pi^\pm$.
% We also make a cut on ${\cal F}$ which keeps
%$67\%$ of the signal and rejects $80\%$ of the background. 
%For Fig.~\ref{fig:mass_hh} and \ref{fig:mass_hpz}, events are 
%sorted by $dE/dx$ according to the most likely hypothesis.
%For Fig.~\ref{fig:mass_hks}, $3\sigma$ consistency with the
%pion hypothesis is required.
%Overlaid on these plots are the projections of the
%PDFs used in the fit, normalized according to the fit results multiplied by 
%the efficiency of the additional cuts ($\sim 50-70\%$ for the signal 
%and $\sim 1-8\%$ for the background).

%/cdat/df/pg/ps/lp99/conttks_prl.ps

%/home/fkw/src/rareb/maxl/run/proj_plots/conthpz99new.ps

%/cdat/df/pg/ps/lp99/kspi_mb_de.eps

%/home/fkw/src/rareb/maxl/run/proj_plots/hpz_mb_kpi_2.eps

To evaluate how systematic uncertainties in the PDFs affect the
statistical significance for modes where we claim first observations,
we repeated the fits for the $h^+h^-$ and $ K^0_S\pi^0$ modes with all
PDFs changed simultaneously within their uncertainties to maximally
reduce the signal yield in the modes of interest. Under these extreme
conditions, the significance of the first-observation modes $\pi^+
\pi^-$ and $K^0_S \pi^0$ becomes $3.2$ and $3.8~\sigma$
respectively. We also evaluate the branching ratios with alternative
analyses using tighter and looser cuts on the continuum
suppressing variable $|\cos\theta_S|$.  These variations correspond
to halving and doubling the background in the fitted sample.  The
branching ratios change under these variations by much less than
the statistical error.

The fitted yields in the remaining modes are not statistically
significant.  We calculate $90\%$\ confidence level (C.L.) upper limit
yields by integrating the likelihood function
\begin{equation}
{\int_0^{N^{UL}} {\cal L}_{\rm max} (N) dN
\over
\int_0^{\infty} {\cal L}_{\rm max} (N) dN}
= 0.90
\nonumber
\end{equation}
where ${\cal L}_{\rm max}(N)$ is the maximum $\cal L$\ at fixed $N$\
to conservatively account for possible correlations among the free
parameters in the fit. We then increase upper limit yields by their
systematic errors and reduce detection efficiencies by their
systematic errors to calculate branching fraction upper limits given
in Table I.

%---------------------------------------------------------------------
%
\section{Information on the Weak Phase $\gamma$}
%
%---------------------------------------------------------------------

%\input{frank.tex}
%As mentioned in the introduction, 
Charmless hadronic $B$ decays are
a sensitive probe of $\gamma$, the phase of the CKM matrix element
$V_{ub}^*$, due to the interference of tree (Fig.~\ref{fig:feynman}(a))
and penguin (Fig.~\ref{fig:feynman}(b)) diagrams.
A number of methods for extracting $\gamma$ from these decays have been
proposed, relying on either the construction of amplitude 
triangles\cite{triangles},
or ratios of CP-averaged branching 
fractions\cite{fleischer-mannel}\cite{neubert-rosner}.
While some of these methods have the virtue of being independent of
model assumptions about strong interaction effects\cite{neubert-rosner},
none of them provide useful constraints 
on $\gamma$ given the present level of precision of our data, as they use
only a restricted set of measurements.

Alternatively, one may trade model independence against exhaustive use of
the available data and attempt a model dependent fit to a large number of
measurements. Such a fit is described in detail
in Ref.~\cite{hou-jgs-fkw}. In this section we will briefly describe the
basic ideas, as well as the main results of this work.

%{\bf NEED TO ADD something that references George's 2 papers.}
%Xiao-Gang~He, Wei-Shu~Hou, and Kwei-Chou~Yang recently 
%pointed out~\cite{he-hou-yang} that a number of the CLEO measurements on
As pointed out in a recent paper by He, Hou, and
Yang,\cite{he-hou-yang} a number of CLEO measurements of charmless
hadronic $B$ decays indicate a preferance for $\cos\gamma < 0.0$.
%They arrived at this conclusion by comparing our
This conclusion is based on a comparison of the CP-averaged branching
fraction measurements with theoretical predictions based on the
factorization model.  In this section we report results of a global
fit that quantifies the qualitative observation made in
Ref.~\cite{he-hou-yang}.

%* He-Hou-Yang, hep-ph/9902256: Original gamma quadrant change observation
%   ===>  Phys. Rev. Lett 83, 1100 (1999)
%        (9 August 1999 issue, day of your talk!)
%* Hou-Yang, hep-ph/9908202: an update including VV modes
% (this paper uses LCSR form factors for PP/PV and get even better numbers)

Assuming factorization one may express\cite{CCTY}\cite{ali} the decay
amplitudes in terms of CKM matrix elements, form factors, decay
constants, quark and meson masses, short distance coefficients, etc.
Many of these are known to more than adequate precision.  In addition,
one may relate many of the poorly known quantities assuming unitarity
of the CKM matrix as well as relationships among form factors.  As a
result, only five poorly known parameters are needed to predict CP
averaged branching fractions for many of the CLEO measurements on
charmless hadronic $B$ decays. These parameters are:

\begin{equation}
  \begin{array}{ccc}
	\gamma & = & Arg(V_{ub}^*) \\
        |V_{ub}/V_{cb}| & & \\
         R_{su} & = & {2m_K^2\over (m_b-m_u)(m_s+m_u)}
                      \sim {2m_K^2\over m_b \times m_s} \\
	F^{B\to\pi} & = & B\to\pi\ \mathrm{transitions\ form\ factor}\\ 
        A^{B\rho}_0 & = & B\to\rho\ \mathrm{transitions\ form\ factor}\\ 
  \end{array}
\end{equation}

We then form a $\chi^2$ between the CLEO results~\cite{all-rareb-conf} 
for the final states
$K^\pm\pi^\mp,\ K^\pm\pi^0,\ K^0_s\pi^\pm,\ K^0_s\pi^0,\ \pi^+\pi^-,
 \ \pi^\pm\pi^0,\ \rho^0\pi^\pm,\ \omega\pi^\pm,\ \rho^\pm\pi^\mp,\ 
K^{*\pm}\pi^\mp,\ \omega K^0_s,\ \omega K^\pm,\ \phi K^0_s,\ \phi K^\pm$
and the theoretical predictions for the 
respective CP-averaged branching fractions.
As additional constraint we add $|V_{ub}/V_{cb}| = 0.08\pm 0.02$, thus
arriving at a $15-5$ degree of freedom fit.
We should note that only 9 of the 14 decay modes are unambiguously
observed (i.e. with statistical significances of more than 4 standard 
deviations). For the remaining 5 final states we see excess yields above
background expectations that have statistical significances 
ranging from $\sim 1.5$ to close to $4$ standard deviations. 

Our choice of decay modes to include was dictated by the following
rationale. We consider for inclusion all decay modes with final states
containing two pseudo-scalars, or a pseudo-scalar and a vector meson,
for which we have results presented at this conference.
We then exclude final states with $\eta$ and $\eta^\prime$ due to the
known ambiguities in predicting these decays~\cite{etaprime-paper}.
We furthermore exclude all final states for which the factorization model
makes predictions that are well below the sensitivity of our present data.

Minimizing the resulting $\chi^2$
we find a global minimum 
of $10.3$ for $10$ degrees of freedom. The parameters at
the minimum are:

\begin{equation}
    \begin{array}{ccc}
	\gamma          & = & 113^{+25}_{-23}\ (161)~\mathrm{degrees} \\
        |V_{ub}/V_{cb}| & = & 0.082 \pm 0.017\ (0.062)\\
        R_{su}          & = & 1.7\pm 0.4\ (0.57) \\
        F^{B\to\pi}     & = & 0.27^{+0.05}_{-0.04}\ (0.43) \\
        A_0^{B\to\rho}  & = & 0.52^{+0.16}_{-0.12}\ (0.78) \\
    \end{array}
\end{equation}

The central value for $R_{su}$ corresponds to $m_s(m_b=4.34$GeV$)$ 
of about $66$ MeV.
The numbers in parentheses are the values for
a second local minimum with only marginally larger 
$\chi^2$. These are the only minima found while repeating the fit $200$ times
with random starting points across a large volume in the five dimensional
parameter space.

\begin{figure}[htbp]
\centering
\leavevmode
\epsfxsize=6.5in
\epsffile{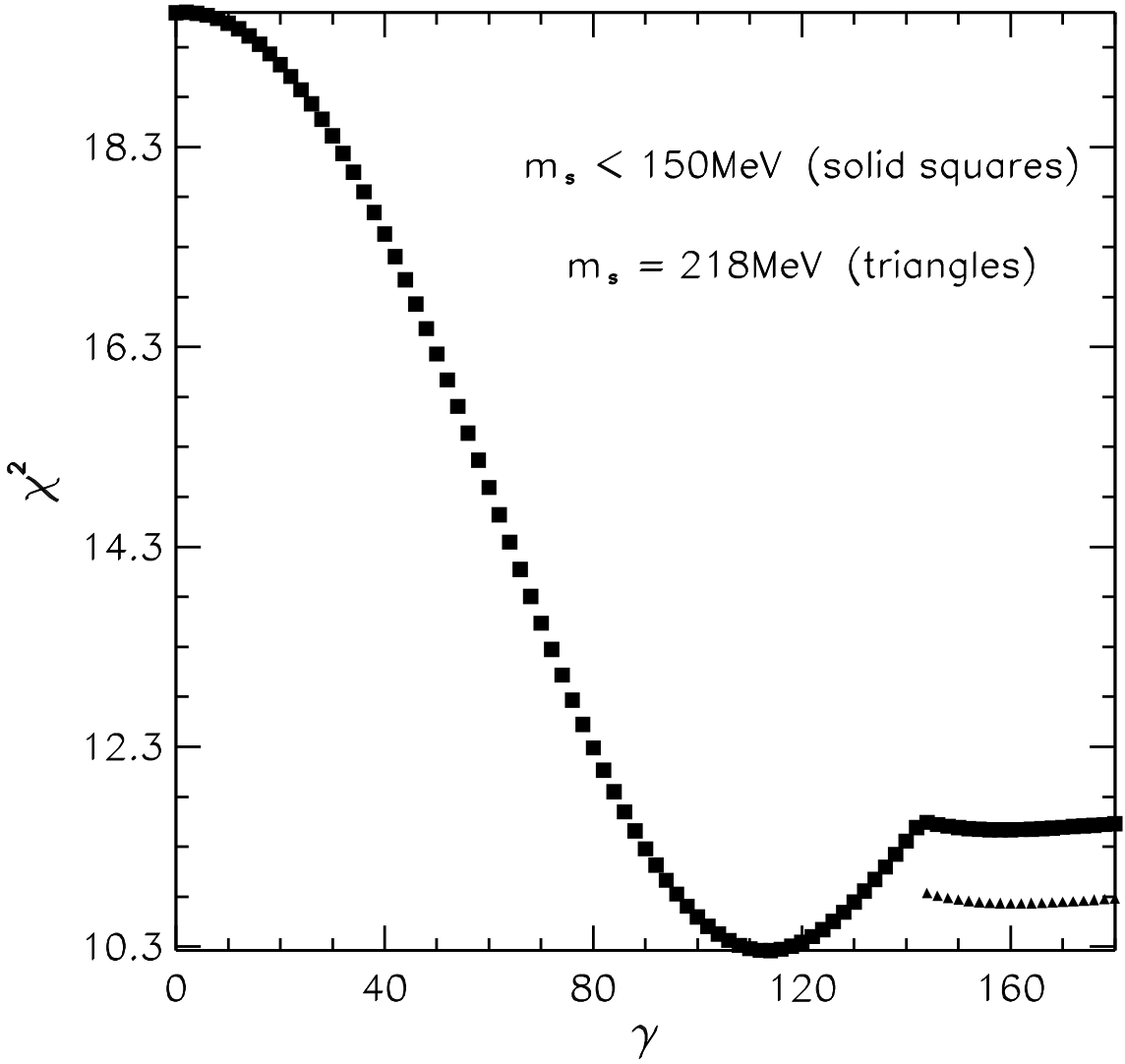}
\caption{$\chi^2$ versus $\gamma$ curve for global fit to 
CP-averaged charmless hadronic $B$ decay branching fractions. 
The triangles depict a local minimum that has the feature of $m_s = 218$MeV.
The squares depict the $\chi^2$ curve with the additional constraint of
$m_s < 150$MeV.}
\label{fig:gamma-chisquare}
\end{figure}

Figure~\ref{fig:gamma-chisquare} shows the dependence of $\chi^2$ 
versus $\gamma$.
We refit at each value of $\gamma$ to allow the fit to find
the global minimum at each fixed value of $\gamma$. This is important as
the correlation matrix for the five free parameters has
significant off-diagonal elements. 
For example, the two most correlated parameters are
$|V_{ub}/V_{cb}|$ and $A_0^{B\to\rho}$ with a correlation coefficient of
$-0.8 $. 
%The effect of this is clearly visible in 
%Figure~\ref{fig:gamma-all}. The four plots show the best fit values for the
%four parameters other than $\gamma$ as a function of $\gamma$ when $\gamma$
%is fixed in the fit.

%To summarize our results from the global fit 
We note that
the best fit values for all parameters are closely consistent
        with theoretical expectations~\cite{CCTY}~\cite{ali}~\cite{LC}.
This is either a surprising coincidence
        or an indication that the model provides an adequate description of
        nature. If we were to believe the latter than we would have to conclude
        that we have made a meaningful measurement of $\gamma$, the phase of
        $V_{ub}^*$.

%---------------------------------------------------------------------
%
\section{Summary}
%
%---------------------------------------------------------------------
In summary, we have measured branching fractions for 
all four exclusive $B\to K\pi$ decays and made first observations of the 
decays $B\to \pi^+\pi^-$ and $B\to K^0\pi^0$.  The latter observation
completes the full set of $B\to K\pi$ measurements. 
In addition, we have shown that a global fit to CP-averaged 
branching fractions measured by CLEO
allows for a model dependent measurement of 
$\gamma = 113^{+25}_{-23}$ degrees.  This is the first
determination of the complex phase of the CKM matrix by any 
method other than the unitarity triangle construction.

%---------------------------------------------------------------------
%
\section{Acknowledgements}
%
%---------------------------------------------------------------------
%Acknowledgements: NEED TO INCLUDE GEORGE HOU!!!!!
We
gratefully acknowledge the effort of the CESR staff in providing us with
excellent luminosity and running conditions.
J.R. Patterson and I.P.J. Shipsey thank the NYI program of the NSF, 
M. Selen thanks the PFF program of the NSF, 
M. Selen and H. Yamamoto thank the OJI program of DOE, 
J.R. Patterson, K. Honscheid, M. Selen and V. Sharma 
thank the A.P. Sloan Foundation, 
M. Selen and V. Sharma thank the Research Corporation, 
F. Blanc thanks the Swiss National Science Foundation, 
and H. Schwarthoff and E. von Toerne thank 
the Alexander von Humboldt Stiftung for support.  
This work was supported by the National Science Foundation, the
U.S. Department of Energy, and the Natural Sciences and Engineering Research 
Council of Canada.

\end{document}